\documentclass[12pt]{iopart}

\usepackage{iopams}
\usepackage{xcolor,soul}
\usepackage[T1]{fontenc}
\usepackage{graphicx}
\sethlcolor{yellow}
\begin{document}

\title[Coherent state transfer  by three-photon STIRAP]{Coherent internal state transfer by three-photon STIRAP-like scheme for many-atom samples}

\author{M. R. Kamsap$^{1,2}$ T. B. Ekogo$^2$, J. Pedregosa-Gutierrez$^1$, G. Hagel$^1$,  M. Houssin$^1$, O. Morizot$^1$, M. Knoop$^1$  and C. Champenois$^1$}

\address{$^1$ Aix-Marseille Universit\'e, CNRS, PIIM, UMR 7345, Centre de Saint J\'er\^ome, Case C21,
13397 Marseille Cedex 20, France \\
$^2$ D\'epartement de Physique, Universit\'e des Sciences et Techniques de Masuku, BP~943, Franceville, Gabon }
\ead{caroline.champenois@univ-amu.fr}
\begin{abstract}
A STIRAP-like scheme is proposed to exploit a three-photon resonance taking place in alkaline-earth-metal ions. This scheme is designed for state transfer between the two fine structure components of the metastable D-state which are two excited states that can serve as optical or THz qu-bit. The advantage of a coherent three-photon process compared to two-photon STIRAP lies in the possibility of exact cancellation of the first order Doppler shift which opens the way for an application  to a sample composed of many ions. The transfer efficiency and its dependence with experimental parameters are analyzed by numerical simulations. This efficiency is shown to reach a fidelity as high as $(1-8.10^{-5})$ with realistic parameters. The scheme is also extended to the synthesis of a linear combination of three stable or metastable states.
\end{abstract}

\maketitle

\section{Introduction}
Atomic coherence has been demonstrated to be an efficient tool for achieving control of the interaction between electromagnetic fields and an atomic sample. It is at the heart of quantum computation techniques based on atomic systems such as cold atoms in cavities or optical lattices as well as strings of trapped ions. For this last system, quantum gates \cite{monroe95, fsk03} and many-qubit entanglement  \cite{haffner05, leibfried05} have been demonstrated, all based on  the building of coherent combinations of internal and/or vibrational states. With trapped ions, two kinds of qubits have been implemented successfully : hyperfine qubits, based on two hyperfine sub-states of the ground state, like in Be$^{+}$, entangled by means of stimulated Raman transitions \cite{monroe95}, and optical qubits, based on two different electronic states, a ground and a metastable one, like the $S_{1/2}$ and $D_{5/2}$ states of Ca$^{+}$,  entangled by means of Rabi pulses \cite{fsk03}. In both cases, the laser pulses are designed to be equivalent to Rabi pulses of exact and well controlled phase. Alternatively, rapid adiabatic passage has been proposed \cite{linington08} and demonstrated its ability to manipulate internal and motional states, in the case of optical qubits \cite{wunderlich07, watanabe11}.

The advantage of adiabatic passage methods is their robustness against technical imperfections of the laser parameters like intensity and phase, which has been studied theoretically \cite{guerin11} and experimentally in the context of trapped ions \cite{poschinger09}. Recent results \cite{noel12} show that robustness compatible with quantum information processing requirements can be reached at the expense of large Rabi frequencies. Nevertheless, these methods remain sensitive to the Doppler effect and the mentioned experimental realizations involve a single ion, cooled to the Doppler limit in  \cite{wunderlich07,noel12} or to the vibrational ground  state in \cite{watanabe11}.

Rapid adiabatic passage is well suited for two level systems; in three level systems, an alternative to Rabi pulses for internal state manipulation   is offered by STIRAP (stimulated Raman adiabatic passage) \cite{oreg84,kuklinski89,laine96,bergmann98} which relies on coherent population trapping (CPT)  in a dark state, made of a linear combination of stable or metastable states.  The paradigmatic system exhibiting such an effect is the $\Lambda$ configuration where two (meta-)stable states are coupled  by light fields to the same excited short-lived state. When the wavelengths of the two involved transitions are very close, for example for hyperfine states or Zeeman sub-levels, the first order Doppler effect is nearly canceled when the two light fields are co-propagating.  This cannot be achieved in a $\Lambda$ scheme  involving three different electronic states. Actually, for very different wavelengths, the contrast reduction of the dark line by Doppler effect can even be used to characterize the motional state of the atom \cite{lisowski05}. For coherent state manipulation, this is a severe drawback and  STIRAP transfer between electronic states was demonstrated on systems larger than single ions only with very close wavelengths \cite{sorensen06}. In that work, STIRAP is driven between the two fine structure terms $D_{3/2}$ and $D_{5/2}$ of the metastable state   of laser cooled Ca$^{+}$ ions, through the $P_{3/2}$ state. Thanks to a relative wavelength difference of 0.5\% and the choice of large one-photon detunings \cite{moller07}, the authors report a transfer efficiency of 90\% with an ion string as well as with a small crystallized cloud of up to 50 ions. Transfer between these two metastable states is of relevance for qubit readout \cite{moller07} and quantum gate processing in the THz domain \cite{toyoda10}. In this last paper, the THz range of the $D_{3/2} \to D_{5/2}$ transition frequency   in Ca$^{+}$ is exploited to phase lock the two laser sources on a passive-type optical comb.  A simple stimulated Raman scheme is then used for completing a Cirac-Zoller gate \cite{cirac95}, but once again, on a single trapped ion cooled to the vibrational ground state \cite{toyoda10}.

To extend these adiabatic passage schemes to many-atom samples, they must be made insensitive to the first order Doppler effect.  This is accomplished if we extend the concept of STIRAP to a three-photon scheme where the  Doppler effect can be exactly canceled by geometric considerations. This is  possible with the  three-photon CPT identified and analyzed in \cite{champenois06}.  This CPT gives rise to a dark line in the fluorescence spectrum which can be made very narrow  and may be used as a THz frequency standard in an ion cloud \cite{champenois07}. In the scheme described in this manuscript, we propose to take advantage of the trapping in a coherent superposition of $S_{1/2}$, $D_{3/2}$ and $D_{5/2}$ state to transfer efficiently the atomic state between the  $D_{3/2}$ and $D_{5/2}$ levels and even create any desired combination of these basic states. Thanks to the exact cancellation of the Doppler effect, the scheme can be applied to a large sample with many  ions, provided that the available laser power is sufficient to reach the required laser coupling for the complete sample. This is a major advantage compared to pulse sequence proposed in \cite{malinovsky97,sola99} to transfer population by STIRAP inspired method in multilevel systems. In these works, alternating STIRAP schemes are compared to straddling STIRAP schemes to transfer population along a chain-wise coupling scheme. In both cases, the study assumes that all the couplings are resonant and there is no apparent resonance involving more than two photons which could lead to a cancellation of the Doppler effect.

The present article is organized as follows. In section~2, the concept of coherent population trapping by three-photon resonance is introduced with a focus on the main results useful for our demonstration. For a full understanding of this coherent process readers are referred to \cite{champenois06}. In section~3, the efficiency of the three-photon STIRAP is analyzed through numerical simulations of the internal state evolution under pulsed laser couplings. The extension of this method to the synthesis of a three-state linear combination is presented in section~4.

\section{Coherent population trapping by three-photon resonance}\label{sec_CPT}
\begin{figure}[htb] \begin{center}
\includegraphics[height=6.cm]{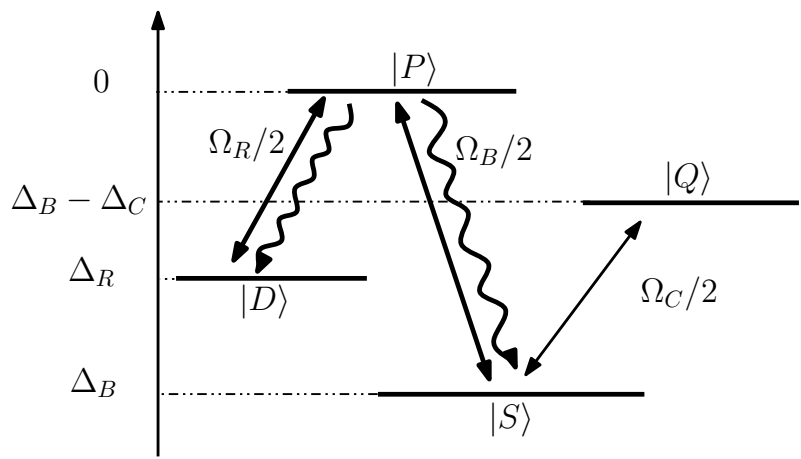}
\caption{$N$-level scheme in the dressed state picture: The states $|D\rangle$,
$|P\rangle$, $|S\rangle$ coupled by laser couplings $\Omega_R$ and $\Omega_B$
form a $\Lambda$-configuration, state
$|S\rangle$ couples weakly to the metastable state $|Q\rangle$ by $\Omega_C$.
The wavy lines indicate the
radiative decay. Parameters and possible atomic species are
discussed in the text. \label{fig_N}} \end{center}
 \end{figure}

The scheme we propose can be applied to any atomic system  composed of four electronic levels
which are coupled by laser fields, according to the $N$-shaped
scheme depicted in Fig.~\ref{fig_N} and where states $|S\rangle$,
$|D\rangle$ and $|Q\rangle$ are (meta)stable while state $|P\rangle$ is
short-lived and decays radiatively into $|S\rangle$ and $|D\rangle$.
This  level configuration  is realized, for instance, in
alkaline-earth atoms with hyperfine structure and in alkaline-earth metal
ions with a metastable $d$-orbital, such as   Ca$^+$,
Sr$^+$, or Ba$^+$. In this manuscript we focus on the ion case  where the
 levels can be identified with the states
$|S\rangle=|S_{1/2}\rangle$, $|P\rangle=|P_{1/2}\rangle$,
$|D\rangle=|D_{3/2}\rangle$, and $|Q\rangle=|D_{5/2}\rangle$. As our experiments are realized using Ca$^+$-ions, we use Ca$^+$ parameters as numerical inputs in the simulations, but the described scheme does not rely on a specific value and transposal to other ions sharing the same internal structure is straightforward. The transitions $|S_{1/2}\rangle \leftrightarrow |P_{1/2}\rangle$ (labeled $B$) and $|D_{3/2}\rangle \leftrightarrow |P_{1/2}\rangle$  (labeled $R$) are electric-dipole allowed and are commonly used for Doppler cooling and detection by induced fluorescence. The coupling scheme also implies the  electric quadrupole transition $|S_{1/2}\rangle \leftrightarrow |D_{5/2}\rangle$ (labeled $C$), which has a spontaneous emission rate of the order of 1 s$^{-1}$ for Ca$^+$. The spontaneous emission rate of the magnetic-dipole transition  $|D_{3/2}\rangle \leftrightarrow |D_{5/2}\rangle$ is of the order of $10^{-6}$ s$^{-1}$ and is neglected.

The three-photon resonance is theoretically introduced  in \cite{champenois06}. It results in a coherent population trapping that  is well explained in the dressed state picture where the non-coupled eigenstates are defined by the hamiltonian 
 \begin{equation}
 H_0 =  \hbar\Delta_R|D\rangle\langle D|+\hbar\Delta_B|S\rangle\langle S|+\hbar(\Delta_B-\Delta_C)|Q\rangle\langle Q|
 \end{equation}
with detunings defined as $\Delta_B=\omega_B-\omega_{PS}$, $\Delta_R=\omega_R-\omega_{PD}$, and
$\Delta_C=\omega_C-\omega_{QS}$. $\omega_X$ is the laser frequency on the $X$ labelled  transition  and $\omega_{IJ}$ is the Bohr frequency of the atomic transition $|I\rangle \leftrightarrow |J\rangle$. $\Omega_{R,B,C}$ are the corresponding Rabi frequencies characterizing the laser couplings. In practice,   the dipole and
the quadrupole couplings differ by a few orders of magnitude and  our description assumes that the $|Q\rangle$ state  is weakly coupled to the
$\Lambda$-scheme formed by the two strong laser couplings involving $|S\rangle$, $|P\rangle$, and $|D\rangle$.
 Considering that $|Q\rangle$ is weakly coupled to   $|S\rangle$, the subsystem ($|S\rangle$,  $|Q\rangle$) can be diagonalized and solved analytically to first order in $\alpha_C=\Omega_C/2\Delta_C \ll 1$. The new eigenstates are then
\begin{equation}
 \left|S_Q\right>  ={\mathcal N} \left(|S\rangle+\alpha_C|Q\rangle\right); \ \ \left|Q_S\right>  ={\mathcal N}\left(|Q\rangle-\alpha_C|S\rangle\right)
\end{equation}
(where ${\mathcal N}$ is the normalization factor) with
eigenfrequencies light-shifted by $\pm \alpha_C\Omega_C/2$.

\begin{figure}[htb] \begin{center}
\includegraphics[height=6.cm]{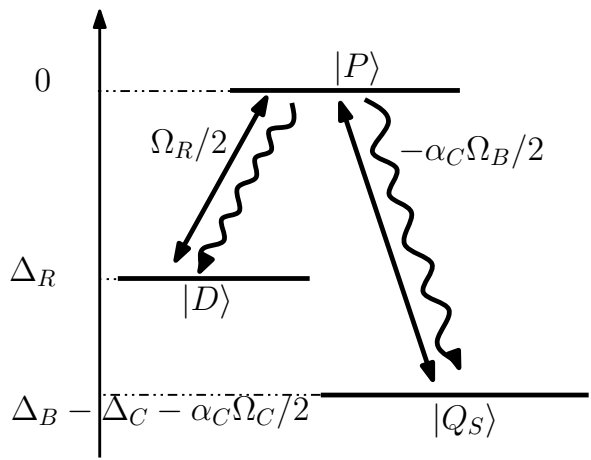}
\caption{Part of the dressed state picture relevant for the three-photon resonance condition, giving rise to the coherent  dark state. The straight lines stand for laser couplings and the wavy lines for radiative decay (see text for details). \label{fig_lambda}} \end{center}
 \end{figure}

 The new eigenstate $\left|Q_S\right>$ is  coupled to
 $\left|P\right>$ by a coupling strength quantified by the Rabi frequency $-\alpha_C\Omega_B$ and the dressed state configuration ends up in a $\Lambda$-scheme based on $\left|Q_S\right>$,$\left|P\right>$ and $\left|D\right>$ state (see Fig.\ref{fig_lambda}).  The radiative processes taken into account
couple $|P\rangle$ to states $|S\rangle$ and $|D\rangle$ by the decay rate $\gamma_P$  and
branching ratio $\beta_{PS}/\beta_{PD}$. $\Lambda$-schemes are well known to give rise to coherent population trapping into a dark state when  the dressed metastable states supporting the $\Lambda$ ( $\left|Q_S\right>$ and $\left|D\right>$ in our case) are degenerated \cite{arimondo96} and are fed by spontaneous emission from the short lived state. The degeneracy condition is fulfilled on  the light-shifted three-photon resonance condition
\begin{equation}\label{3-photon}
\Delta_{\mathit{eff}}=\Delta_R+\Delta_C-\Delta_B+\alpha_C \Omega_C/2=0.
\end{equation}
The spontaneous emission rate from $|P\rangle$ to $\left|Q_S\right>$ is in second order in $\alpha_C$. For  appropriate $\alpha_C$ values, this rate is sufficient to lead to effective population trapping but sets a minimum time scale boundary for its efficiency.

Provided that the three-photon and two-photon ($\Delta_R=\Delta_B$) resonance conditions are sufficiently split apart ({\it i. e.} $\Delta_C$ is larger than the relevant dark line widths, see \cite{champenois06} for justification), the atomic system is then pumped into the dark state
 \begin{equation}
|\Psi_{D}\rangle ={\cal N}'\left({\cal E}|D\rangle+|Q_S\rangle\right) \label{psiD}
\end{equation}
with ${\cal E}=\alpha_C\Omega_B/\Omega_R$ and normalization factor ${\cal N}'$. The dark state stability is fundamentally limited by the radiative decay of states $|Q\rangle$ and  $|D\rangle$ which is of the order of 1~s for Ca$^+$, 350~ms for Sr$^+$ and more than 10~s for Ba$^+$. The reduction of the dark state lifetime  by the first order Doppler effect   can be cancelled  in the Doppler-free configuration \cite{grynberg76} which is defined by the phase matching condition
\begin{equation}
\Delta {\bf k}={\bf k_R}+{\bf k_C}-{\bf k_B}={\bf 0}.
\end{equation}
where ${\bf k_X}$ is the wave vector of laser $X$. In the following, we assume this matching condition is satisfied and we neglect any Doppler effect including the second order Doppler effect which results in a line broadening and shift which are not relevant here \cite{champenois07}.

Inspired by the original STIRAP method \cite{oreg84,kuklinski89,bergmann98}, we want to coherently transfer atomic population between the two qu-bit states $|D\rangle=|D_{3/2}\rangle$ and $|Q\rangle=|D_{5/2}\rangle$ by adiabatic following of the dark state.  One of the main concerns comes from the definition of the dark state itself (Eq.~\ref{psiD}) which includes a small part of the $|S_{1/2}\rangle$ state. This issue as well as the identification of the relevant parameters  which offers a compromise between a small enough coupling on the $C$ transition and a fast STIRAP are addressed in the following section. Population  transfer along a chain of 4 states coupled in a  $N$-scheme  was already studied  in \cite{oreg92} in the assumption of three-photon resonance. The delayed pulse method analyzed in \cite{oreg92} does not rely on a three-photon dark state but rather on  successive two-photon Raman transfers and the process remains sensitive to the Doppler effect.

\section{STIRAP with Gaussian pulses}
 Numerical simulations of the internal dynamics of the atomic state are made by integrating the master equation for the density matrix resulting from the optical Bloch equations. We first consider that the weak coupling on the quadrupole electric transition is always on and is kept constant. The laser detunings are fixed all along the state transfer and, in a first step, we assume that the three involved lasers are phase-locked to the same frequency comb such that their relative phase drift is negligible. The laser intensities on the $B$ and $R$ transitions are modulated in order to have a complete  overlap between the dark state and the desired atomic state at the beginning and end of the transfer. The commonly used Gaussian shape for STIRAP laser pulses are not the optimal choice and can be optimized to reach better fidelity by minimization of non-adiabatic losses \cite{vasilev09}. Nevertheless, to focus on the atomic system response and compare to previous theoretical \cite{moller07} and experimental works \cite{sorensen06}, we keep a simple Gaussian profile for the time dependence of the pulses, which is a relevant representation of what can be experimentally produced using first order diffraction from acousto-optical modulators  :
 \begin{eqnarray}\label{eq_pulses}
\Omega_B(t)&=&\Omega_B^0 \exp\left[-\left(\frac{t\pm\Delta t/2}{\tau}\right)^2\right],  \nonumber \\
\Omega_R(t)&=&\Omega_R^0 \exp\left[-\left(\frac{t\mp\Delta t/2}{\tau}\right)^2\right].
\end{eqnarray}
The width of the laser pulses is defined by $\tau$ and $\Delta t $ is their time delay. The order of application of the pulses (the $\pm$ sign) depends on the desired transfer and follows the non-intuitive STIRAP requirement : the first choice drives a transfer from $D_{3/2}$ to  $D_{5/2}$, the second choice from $D_{5/2}$ to  $D_{3/2}$. To follow the internal dynamics, we solve the master equation for the density matrix $\rho$
\begin{equation} \label{Master:Eq}
\frac{\partial}{\partial t}\rho=-\frac{\rm i}{\hbar}[H,\rho]+{\cal L}\rho \end{equation}
with $H=H_0+H_I(t)$. $H_I(t)$ includes  the laser couplings  by
 \begin{equation} \label{Ham}
H_I(t) = \frac{\hbar\Omega_B(t)}{2}|P\rangle\langle S|+\frac{\hbar\Omega_R(t)}{2}|P\rangle\langle D| +\frac{\hbar\Omega_C}{2}|Q\rangle\langle S|+{\rm H.c.}
\end{equation}
and the relaxation operator
\begin{eqnarray}
{\cal L}\rho
&=&-\frac{1}{2}\gamma_P \left(\rho|P\rangle\langle
P|+|P\rangle\langle P|\rho\right) \label{eq_L}\\
& &+\beta_{PS}\gamma_P|S\rangle\langle P|\rho|P\rangle\langle S|
+\beta_{PD}\gamma_P|D\rangle\langle P|\rho|P\rangle\langle D|  \nonumber
 \end{eqnarray}
 describes the radiative processes,
coupling $|P\rangle$ to states $|S\rangle$ and $|D\rangle$, with a
branching ratio $\beta_{PS}/\beta_{PD}= 14.4$ for Ca$^{+}$ \cite{safronova11} and
$\beta_{PS}+\beta_{PD}=1$. The $P_{1/2}$ state lifetime $\gamma_P^{-1}$  has been measured to be $7.07 \pm 0.07$~ns in \cite{gosselin88} and $7.1 \pm 0.02$~ns in \cite{jin93}. Very precise calculations \cite{safronova11} recommend to use $\gamma_P^{-1}=6.87 \pm 0.13$~ns. For the simulations, we use $\gamma_P^{-1}=7.00$~ns as a numerical value but the accuracy of the parameter is not relevant for the STIRAP process.

\subsection{Transfer efficiency}
We start with an atom in the $D_{3/2} = |D\rangle$ state and the target  state of the STIRAP-like transfer is expected to be $|Q_S\rangle  ={\mathcal N}(|Q\rangle-\alpha_C|S\rangle) $, if the approximations used in \cite{champenois06} are still valid. The fidelity of the transfer is then
\begin{equation}
F=\langle Q_S |\rho|Q_S\rangle=(\alpha_C^2\rho_{SS}+(1-\alpha_C^2)\rho_{QQ}-2\alpha_C {\rm Re}(\rho_{SQ}))
\end{equation}
in second order in $\alpha_C$. Our final objective is a complete transfer from $D_{3/2}$ to $D_{5/2}$. Once the STIRAP process completed, the weak coupling laser has to be switched off to reduce the contribution of $|S\rangle$ to zero. This can be done with an exponential decay of $\Omega_C$ which can be made as short as 1~$\mu$s without any coherence loss. In this case, the target state is simply $|Q\rangle$ and the transfer efficiency is quantified by the average occupation probability $P_Q=\rho_{QQ}$. Let's mention that the exact delay between the end of the STIRAP and the decay of the weak coupling does not have to be controlled with a high precision as the intermediate state $|Q_S\rangle$ is stable on time scales of the order of the 1~s lifetime.

\begin{figure}[htb]
   \begin{center}
\includegraphics[height=6.cm]{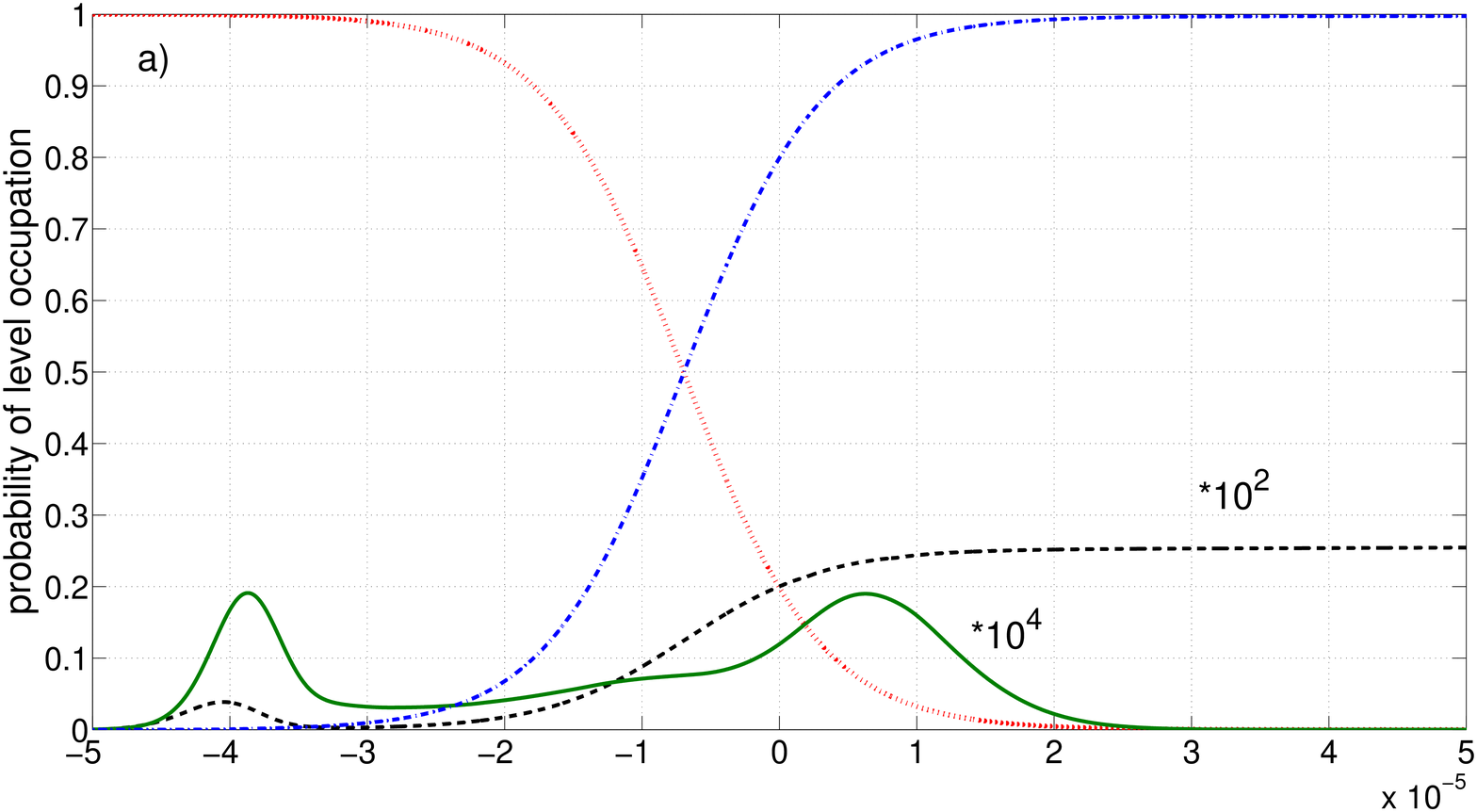}\\
\includegraphics[height=6.cm]{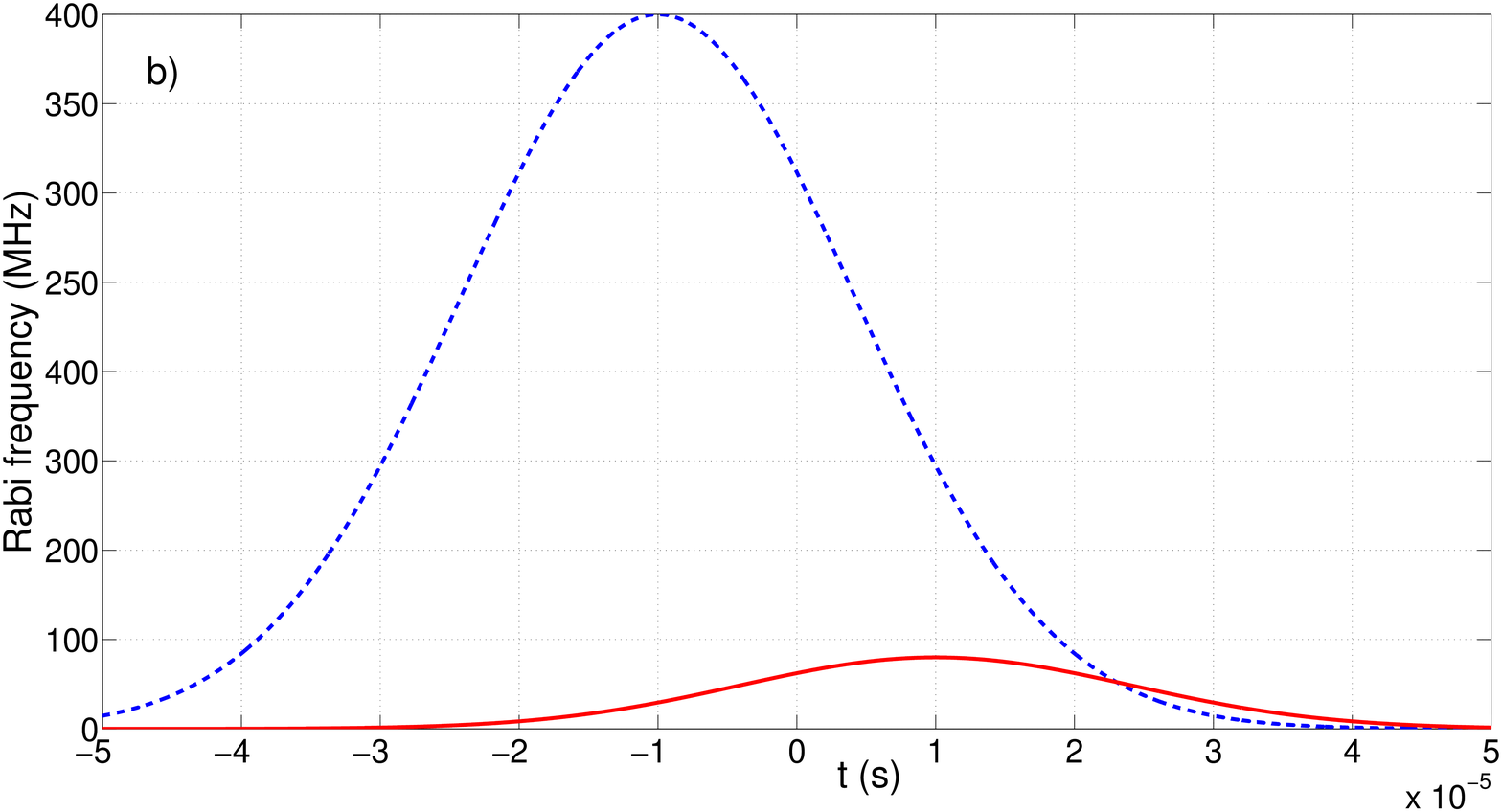}
\caption{ a): Time evolution of the population of the $D_{3/2}$ (red dotted line), $P_{1/2}$ ($\times 10^4$, green solid line), $S_{1/2}$ ($\times 10^2$, black dashed line)and $D_{5/2}$ (dot-dashed blue line) states during the STIRAP process driven by the Gaussian pulses $\Omega_B(t)$ and $\Omega_R(t)$ (see Eq.~\ref{eq_pulses}). Laser parameters are $\tau=\Delta t=20 \mu$s, $\Omega_C/2\pi=10$~MHz, $\Delta_C/2\pi=100$~MHz, $\Omega_B^0/2\pi=400$~MHz, $\Delta_B/2\pi=100$~MHz, $\Omega_R^0/2\pi=40$~MHz,  $\Delta_R=\Delta_B-\Delta_C-\alpha_C \Omega_C/2$.   b): Time evolution of the Rabi frequency $\Omega_B(t)$ (blue dashed line) and $\Omega_R(t)$ (red solid line) $\Omega_C$ is constant during the STIRAP process.. \label{fig_DtoQS}}
\end{center}
 \end{figure}

Following the numerical analysis of a regular two-photon STIRAP process in Ca$^{+}$ \cite{moller07}, we want to define a process where the two branches of the $\Lambda$-scheme  see a maximum coupling strength of the same order of magnitude. This implies that  $\Omega_B^0$ must be large enough to compensate for the weak mixing term $\alpha_C \ll1$.  Indeed, comparison of numerical results shows that very good transfer efficiencies are observed for a mixing parameter $\alpha_C$ close to 0.05 and all the results presented in this paper were obtained with $\alpha_C=0.05$.  In practice, it implies that $\Omega_B^0/2\pi$ must be of the order of a few hundreds of MHz, which is certainly a strong experimental constraint, even more when the transfer is addressed to an ion cloud rather than a single ion. Concerning the optimal delay between the two pulses, our simulations confirm the property demonstrated in \cite{gaubatz88} that optimum transfer is observed for a delay $\Delta t$ equal to $\tau$, which is close to the half-width of the pulse, with our notation (Eq.~\ref{eq_pulses}). Figure~\ref{fig_DtoQS} shows the time evolution of the population of the atomic  states along a population transfer process from $D_{3/2}$ to $D_{5/2}$. We assume that the atomic system is previously prepared in the $D_{3/2}$ state, which is easy to realize by keeping only the $B$-laser on, without any repumping out of $D_{3/2}$ (see level scheme on Fig.~\ref{fig_N}).
The laser parameters used for the simulation are given in the figure's caption. The fidelity  $F$ reached at the end of the STIRAP and  $P_Q$ at the end of the total transfer both equal $(1-8.10^{-5})$. For the chosen parameters, the experimental duration must be 100~$\mu$s  in order  to reach such an excellent fidelity.  This duration can be reduced by a factor of 2 (50~$\mu$s) if the required fidelity is $(1-3.10^{-4})$ and by a factor of 4 (27~$\mu$s) if a fidelity of $(1-1.10^{-3})$ is sufficient.

In order to reverse the transfer, starting in the $D_{5/2}$ state, this state must first be dressed and coupled to become $|Q_S\rangle$. This can be achieved by adiabatic rapid passage with two alternatives : either in the no-crossing case where the detuning is kept fixed and the laser coupling is increased to reach the desired dressed state, or in the crossing case where the laser coupling is fixed and the detuning is chirped from a very large value to the target one (see \cite{vitanov01} for a complete review). In our simulations, we used the first method and could bring the atomic state to $|Q_S\rangle$ by switching on the weak coupling on a time scale of the order of 1~$\mu$s, with a fidelity reaching 1 to better than $10^{-6}$. Then, the STIRAP pulses are applied in the reversed order (in practice, with the other set of signs in Eq.~\ref{eq_pulses}), to adiabatically follow the dark state to $D_{3/2}$.

\begin{figure}[htb]
   \begin{center}
\includegraphics[height=6.cm]{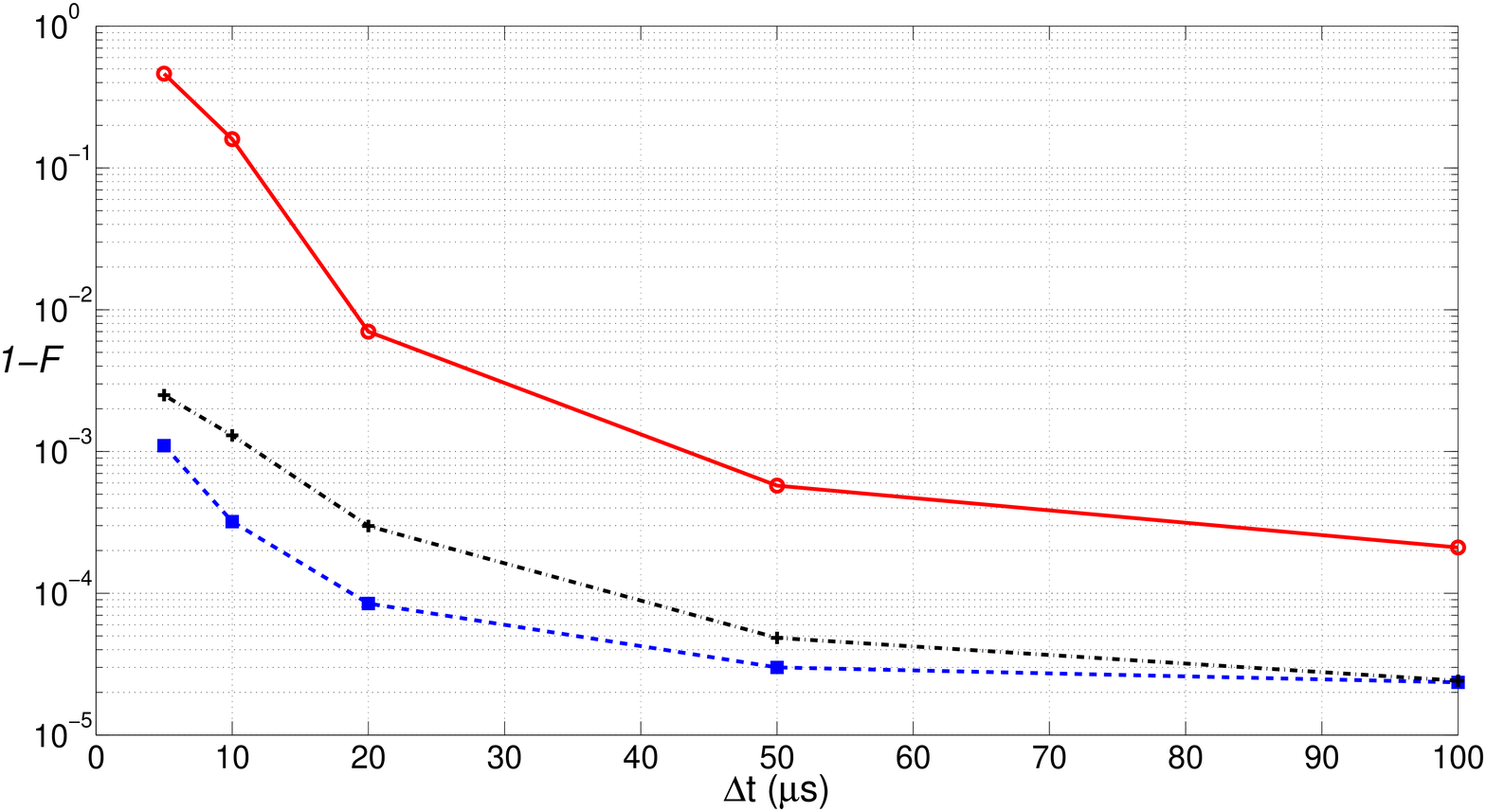}
\caption{Non-fidelity $1-F=1-P_Q$  of the full transfer driven by Gaussian pulses plus weak coupling decay versus pulse duration and delay $\tau=\Delta t$ (see Eq.~\ref{eq_pulses})  Laser parameters are   $\Omega_B^0/2\pi=400$~MHz, $\Delta_B/2\pi=100$~MHz, $\Omega_R^0/2\pi=40$~MHz and $\Delta_R=\Delta_B-\Delta_C-\alpha_C \Omega_C/2$. The open red circles on the solid line are for $\Omega_C/2\pi=1$~MHz, $\Delta_C/2\pi=10$~MHz, the black crosses on the dash-dotted line are for $\Omega_C/2\pi=5$~MHz, $\Delta_C/2\pi=50$~MHz and the full blue squares on the dashed line is for $\Omega_C/2\pi=10$~MHz, $\Delta_C/2\pi=100$~MHz. \label{fig_Fdt}}
\end{center}
 \end{figure}

\begin{figure}[htb]
   \begin{center}
\includegraphics[height=6.cm]{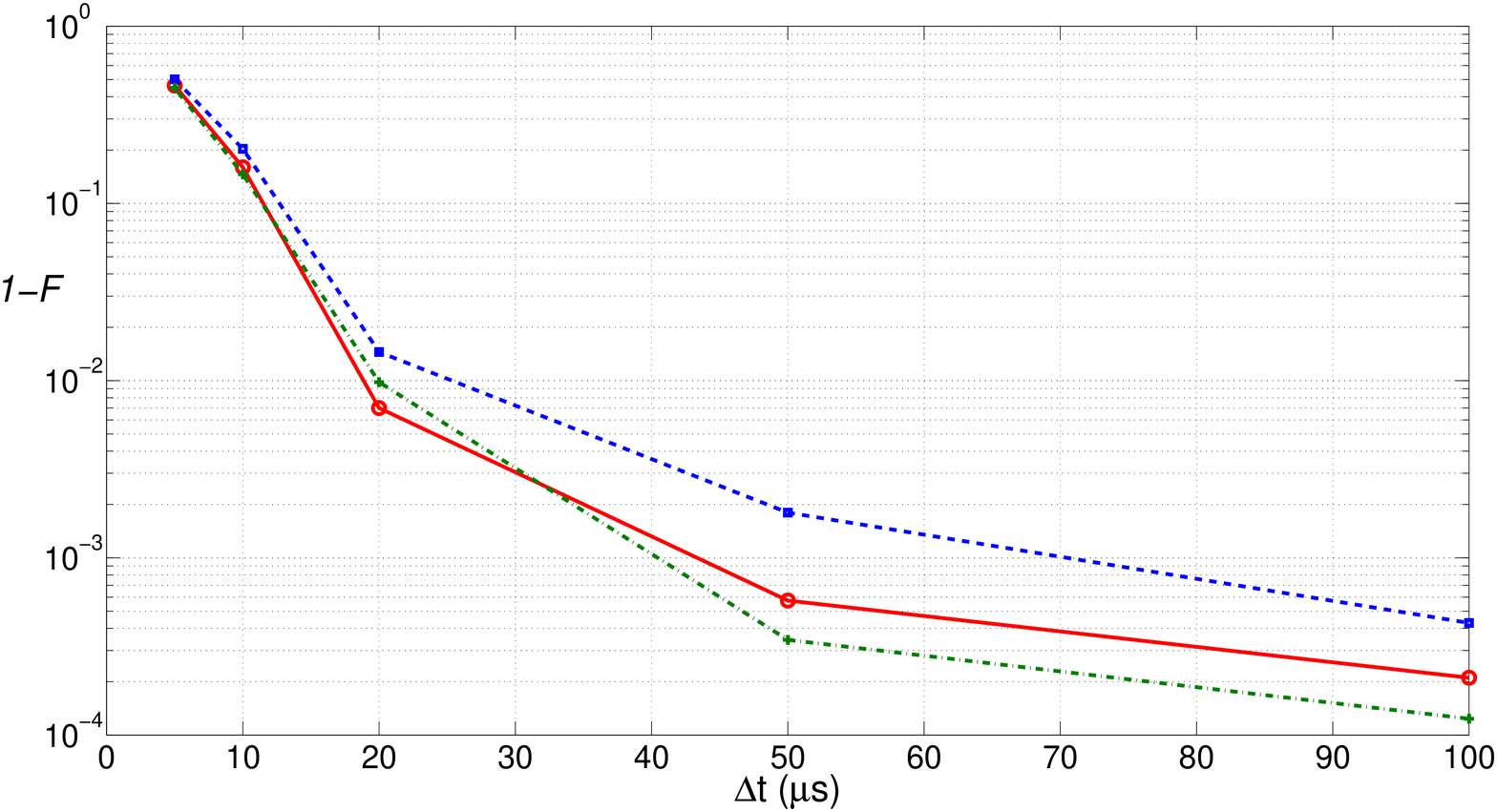}
\caption{Non-fidelity $1-F=1-P_Q$ of the full transfer driven by Gaussian pulses plus weak coupling decay versus their duration and delay $\tau=\Delta t$ (see Eq.~\ref{eq_pulses}).  Laser parameters are  $\Omega_C/2\pi=1$~MHz, $\Delta_C/2\pi=10$~MHz, $\Omega_B^0/2\pi=200$~MHz and $\Omega_R^0/2\pi=20$~MHz (filled square, blue dashed line), $\Omega_B^0/2\pi=400$~MHz and $\Omega_R^0/2\pi=40$~MHz (empty circle, red solid line),  $\Omega_B^0/2\pi=800$~MHz and $\Omega_R^0/2\pi=80$~MHz (cross, green dot-dashed line), $\Delta_B/2\pi=100$~MHz, and $\Delta_R=\Delta_B-\Delta_C-\alpha_C \Omega_C/2$.  \label{fig_FdtObOr}}
\end{center}
 \end{figure}

This fast and complete transfer is possible at the expense of a large coupling on the weak transition. Figure~\ref{fig_Fdt} shows how the evolution of the fidelity with  the characteristic pulse time depends on the weak coupling $\Omega_C$ but keeping the coupling parameter $\alpha_C=0.05$ constant.
The comparison of the fidelity evolution for different sets of ($\Omega_C$, $\Delta_C$) clearly shows that this transfer can reach a very good fidelity if sufficient laser power is available. It takes of the order of 5~mW focused on a beam radius of 10~$\mu$m to reach a Rabi frequency of 1~MHz on the quadrupole transition of Ca$^{+}$. With the progress made in coherent and powerful lasers, achieving a 10~MHz coupling strength with a larger beam size is not out of reach \cite{bolpasi2012}. Another strategy could be offered by increasing the coupling strength on the two main transitions $B$ and $R$. Figure~\ref{fig_FdtObOr} shows that increasing $\Omega_B(t)$ and $\Omega_R(t)$ results in a better fidelity, but this gain can be seen only for the longer  pulses and its magnitude is not sufficient  to compensate for a too small coupling on the weak transition.

\subsection{Sensitivity to experimental imperfections}
\begin{figure}[htb]
   \begin{center}
\includegraphics[height=6.cm]{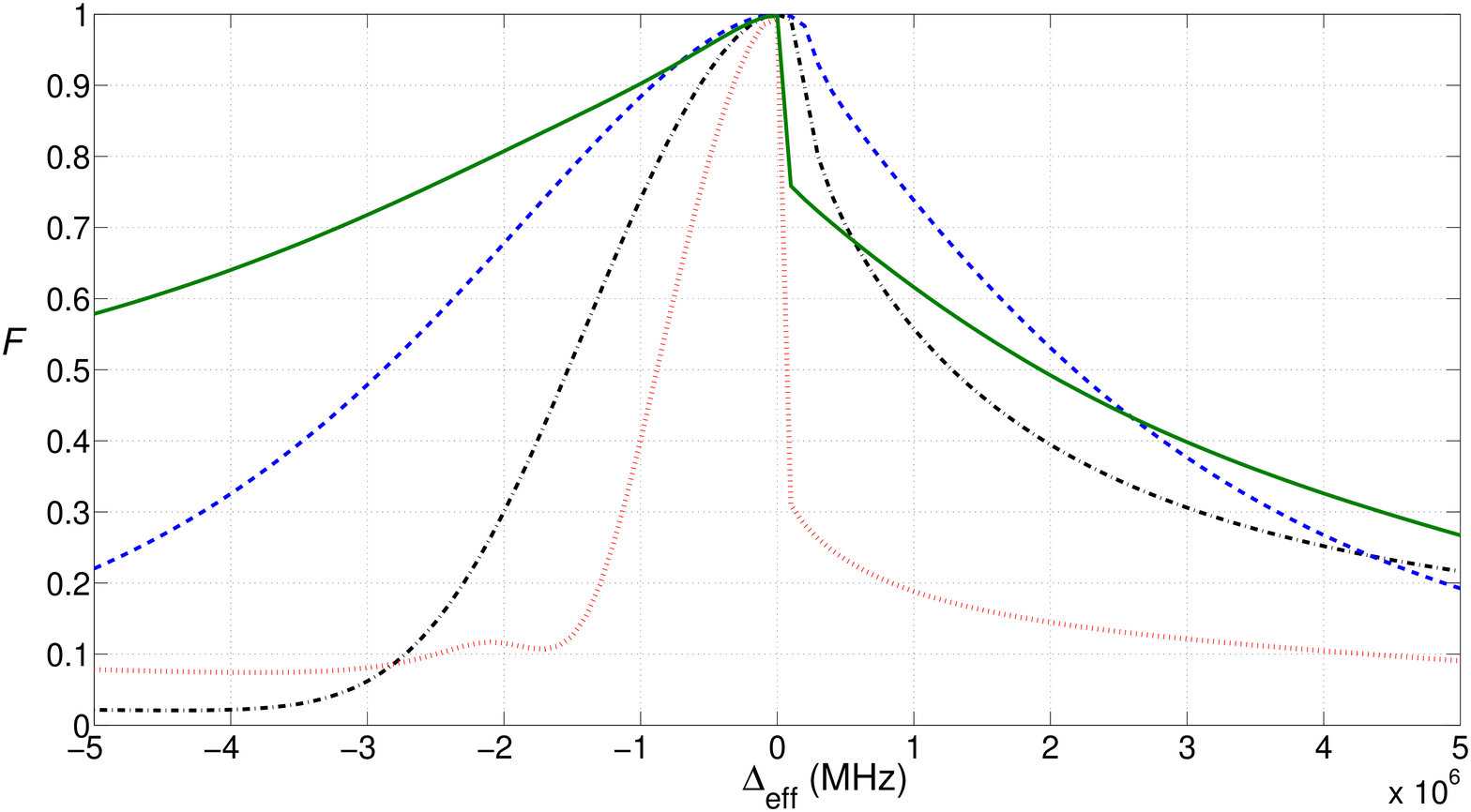}
\caption{Fidelity $F$ of the full transfer driven by Gaussian pulses plus weak coupling decay versus the three-photon resonance mismatch $\Delta_{\mathit{eff}}=\Delta_R-(\Delta_B-\Delta_C-\alpha_C \Omega_C/2)$. Common laser parameters are  $\Omega_B^0/2\pi=400$~MHz and $\Omega_R^0/2\pi=40$~MHz. The strong coupling cases for  the weak  transition are  for  $\Omega_C/2\pi=10$~MHz, $\Delta_C/2\pi=100$~MHz, $\Delta t=20~\mu$s and $\Delta_B/2\pi=100$~MHz (dashed blue line) and  $\Delta_B/2\pi=190$~MHz (dot-dashed black line).  The weak coupling cases are for $\Omega_C/2\pi=1$~MHz, $\Delta_C/2\pi=10$~MHz $\Delta t=45~\mu$s, $\Delta_B/2\pi=100$~MHz (red dotted line) and $\Delta_B/2\pi=10$~MHz (green solid line)\label{fig_FdeltaR}}
\end{center}
 \end{figure}

To be satisfied, the three-photon resonance condition requires the control of the  relative detuning of three lasers. The relative detuning relation given in Eq.~\ref{3-photon} includes a light-shift that can be rewritten like $\alpha_C^2 \Delta_C$. With the identified optimum condition $\alpha_C=0.05$, it means that this light-shift is equal to 0.25\% of the weak coupling detuning. This is very small compared to the effective line-width of the STIRAP efficiency shown on figure~\ref{fig_FdeltaR} for two different sets of laser parameters: the weak coupling case ($\Omega_C/2\pi=1$~MHz, $\Delta_C/2\pi=10$~MHz) with a pulse delay $\Delta t=45~\mu$s chosen to reach a fidelity better than $(1-10^{-3})$ for no mismatch and the strong coupling case ($\Omega_C/2\pi=10$~MHz, $\Delta_C/2\pi=100$~MHz) with a pulse delay $\Delta t=20~\mu$s chosen to reach a fidelity better than $(1-10^{-4})$ for no mismatch.  Comparison of several curves of this figure confirms that the sensitivity of transfer efficiency to detuning mismatch is strongly controlled by the one photon detuning $\Delta_R\simeq \Delta_B-\Delta_C$, like already observed for two-photon STIRAP in \cite{moller07}. This is clearly illustrated when comparing the red dotted line and the green solid line, for the weak coupling case or when comparing the dashed blue line and the dot-dash black line for the strong coupling case. For each couple the difference in laser parameters lies in the one photon detuning which is nearly null for the broader curve and equal to 90~MHz for the narrower ones. So a smaller one-photon detuning makes the STIRAP efficiency less sensitive to detuning mismatch. For a given one-photon detuning, the curve is also made broader by a stronger coupling on the weak transition. The asymmetry of the line profile is due to non-adiabatic crossing with a fast decaying state, like already identified in \cite{moller07}.  The profile is symmetric and larger if the time allowed for the transfer is extended. In the strong coupling case with zero one-photon detuning, a detuning mismatch of $\pm 0.1$~MHz leads to a reduced fidelity of 0.997 for a pulse delay of 20~$\mu$s.

\begin{figure}[htb]
   \begin{center}
\includegraphics[height=6.cm]{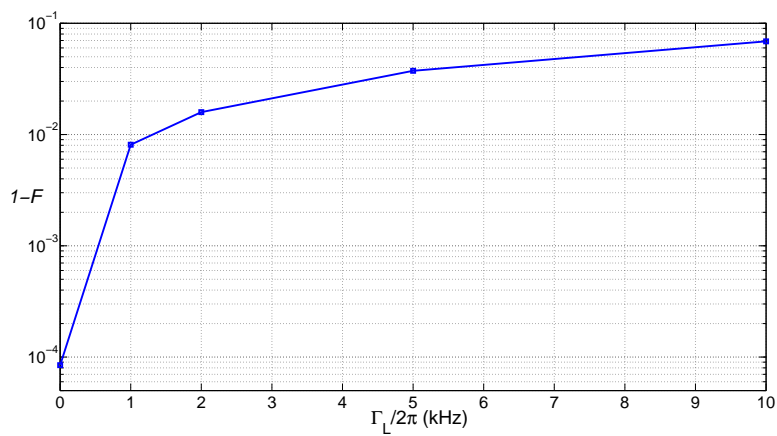}
\caption{Non-fidelity $1-F$ of the full transfer driven by Gaussian pulses ($\Delta t=28~\mu$s) plus weak coupling decay versus the laser half linewidth at half maximum $\Gamma_L$. Laser parameters are  $\Omega_B^0/2\pi=400$~MHz and $\Omega_R^0/2\pi=40$~MHz  and  $\Delta_B/2\pi=100$~MHz, $\Omega_C/2\pi=10$~MHz, $\Delta_C/2\pi=100$~MHz  and $\Delta_R=\Delta_B-\Delta_C-\alpha_C \Omega_C/2$.\label{fig_FGL}}
\end{center}
 \end{figure}

A major experimental imperfection comes from the coherence loss induced by the phase fluctuations of the laser fields which induce a phase fluctuation in the definition of the dark state. It is possible to take into account these phase fluctuations in the relaxation operator of the master equation (Eq.~\ref{eq_L}) by an average line-width responsible for coherence decay \cite{cct_houches75}. It is introduced in the master equation by a Lindblad operator
\begin{equation}
{\cal L}_{relax}\rho=-\frac{1}{2}\sum_m C_m^{\dagger}C_m\rho+\rho C_m^{\dagger}C_m+\sum_m C_m \rho C_m^{\dagger}
\end{equation}
with a $C_m$ operator associated to each laser coupling~\cite{molmer93,PC_JE}:
\begin{eqnarray}
C_m^B &=& \frac{\sqrt{\Gamma_L^B}}{2}\left(|P\rangle\langle P|+|D\rangle\langle D|-|S\rangle\langle S|-|Q\rangle\langle Q|\right) \nonumber \\
C_m^R &=& \frac{\sqrt{\Gamma_L^R}}{2}\left(|D\rangle\langle D|-|P\rangle\langle P|-|S\rangle\langle S|-|Q\rangle\langle Q|\right)  \\
C_m^C &=& \frac{\sqrt{\Gamma_L^C}}{2}\left(|Q\rangle\langle Q|-|S\rangle\langle S|-|P\rangle\langle P|-|D\rangle\langle D|\right) \nonumber
\end{eqnarray}
with $\Gamma_L^X$ the half-width at half-maximum of the spectral width of the laser on the $X$-transition. On Figure \ref{fig_FGL}, we show the evolution of the transfer efficiency when the three lasers have the same line-width.  The results of the simulation show that this efficiency is very sensitive to the laser frequency fluctuations and already with $\Gamma_L=1$~kHz, the fidelity decreases to $(1-2.10^{-2})$ whereas it is larger than $(1-10^{-4})$ for no frequency fluctuations. To take advantage of the three-photon resonance, it is then highly relevant to lock the three lasers on a very stable reference in order to reduce the relative frequency drift. This can be done by locking each laser's frequency on a peak of a frequency comb \cite{reichert99}. For a free running optical comb and the Ca$^{+}$ transition wavelengths ($B: 397$~nm, $C: 729$~nm, $R: 866$~nm) the relative frequency drift is of the order of 1~kHz/s \cite{PC_AAK}. It can be reduced to 40~Hz/s for an optical comb locked on an RF reference and to 1~Hz/s if the reference is in the optical domain \cite{PC_AAK}. With the RF reference, the fidelity is only reduced to $(1-5.10^{-4})$ but with such good performances, other frequency fluctuations like the one induced by the Zeeman shift may surpass the laser line-width effect.

 \section{Building a linear combination of (meta-)stable states}
\begin{figure}[htb]
   \begin{center}
  \includegraphics[height=6.cm]{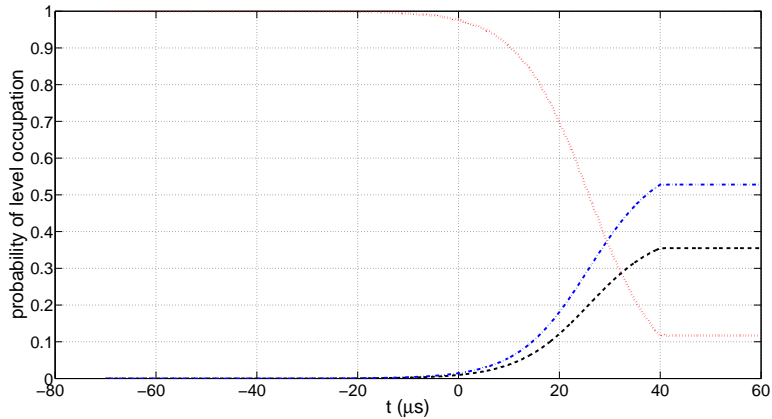}
\caption{Time evolution of the population of the $D_{3/2}$ (red dotted line), $S_{1/2}$ (black dashed line) and $D_{5/2}$ (dot-dashed blue line) states during an incomplete STIRAP process driven by  Gaussian pulses $\Omega_B(t)$ and $\Omega_R(t)$ kept constant since $t=40~\mu$s (see Eq.~\ref{eq_pulses}). Laser parameters are $\tau=\Delta t=28 \mu$s, $\Omega_C/2\pi=50$~MHz, $\Delta_C/2\pi=10$~MHz, $\Omega_B^0/2\pi=400$~MHz, $\Delta_B/2\pi=100$~MHz, $\Omega_R^0/2\pi=40$~MHz,  $\Delta_R=\Delta_B-\Delta_C(1+\sqrt{1+4\alpha_C^2})/2$. \label{fig_DtoQSD}}
\end{center}
 \end{figure}

 As the dark state is formally built out of the $|D\rangle, |Q\rangle, |S\rangle$ dressed states, it is possible to build a linear combination of these states by controlling their contribution  by the laser coupling ratio.  To be more precise, the dark state is a combination of $|D\rangle$ and $ |Q_S\rangle$ and by controlling the  ratio ${\cal E}$ (see Eq.~\ref{psiD}), one can build any desired combination of these two states.  Figure~\ref{fig_DtoQSD} shows the evolution of the population of the dressed states along a pulsed STIRAP, interrupted before completion. From this interruption, all the laser couplings are kept constant as they are. For the demonstration, we choose a non-negligible contribution of the $|S\rangle$ state, which requires to go beyond the weak-coupling approach used in section~\ref{sec_CPT}. The exact solutions for the eigenstates of the $(|Q\rangle, |S\rangle)$ coupled system are well known and  the more general form for $|Q_S\rangle$ can be expressed with the new parameter $\alpha=2\alpha_C/(1+\sqrt{1+4\alpha_C^2})$ like
\begin{equation}
 \left|Q_S\right>  =\frac{1}{\sqrt{1+\alpha^2}}|Q\rangle-\frac{\alpha}{\sqrt{1+\alpha^2}}|S\rangle.
\end{equation}
  Its eigen-energy is $\lambda_Q=-\Delta_C(1+\sqrt{1+4\alpha_C^2})/2$ and this dressed state is coupled to the $|P\rangle$ state by the laser coupling $-\beta \Omega_B/2$ with $\beta=\alpha/\sqrt{1+\alpha^2}$.
The three-photon resonance condition becomes
\begin{equation}\label{3-photon_bis}
\Delta_R-\Delta_B+\Delta_C(1+\sqrt{1+4\alpha_C^2})/2=0.
\end{equation}
 and the laser parameters for figure~\ref{fig_DtoQSD} obey this condition. For continuity reason with the previous section, we build the dark state with the $|Q_S\rangle$ state, which implies that the contribution of $|S\rangle$ can not exceed the one from $|Q\rangle$. If the reverse situation is required, the dark state must be built with $|S_Q\rangle$ by adapting the three-photon resonance condition to the other eigen-energy $\lambda_S=-\Delta_C(1-\sqrt{1+4\alpha_C^2})/2$. The fidelity of the process is evaluated by computing $\langle \Psi |\rho|\Psi \rangle$ with $|\Psi \rangle={\cal M}({\cal E^{'}}|D\rangle+|Q_S\rangle)$, ${\cal M}$ being a normalization factor  and ${\cal E^{'}}=\beta\Omega_B/\Omega_R$. All along the population transfer depicted on figure~\ref{fig_DtoQSD}, the fidelity remains higher than $(1-10^{-4})$ and reaches $(1-2.10^{-5})$ at the end of the process. This very good value shows that the atomic state exactly follows the dark state, even if it is now extended to large couplings on the weak transition. The large value chosen for $\Omega_C$ (50~MHz) in this calculation may not be experimentally realistic but this choice was  made to demonstrate the validity of the description in a broad range of parameters.

 Once the combination built, the $B$- and $R$- coupling lasers can be turned off simultaneously, with a 1~$\mu$s  exponential decay, and leave the system in the equivalent combination of $D_{3/2}$ and $ |Q_S\rangle$. Obviously, if the $C$- coupling laser is also turned off, the $ |Q_S\rangle$ state continuously tends to $D_{5/2}$ (or to $S_{1/2}$ if the dark state is built with $ |S_Q\rangle$). A three-state linear combination therefore requires at least one coupling laser to exist.

 \section*{Conclusion}
The dark state, made of three stable or metastable states and resulting from a three-photon dark resonance is used to coherently transfer population between the two fine-structure states $D_{3/2}$ and $D_{5/2}$ of calcium-like ions. In the ideal case of exact cancelling of the Doppler effect and a phase lock of the three involved lasers, fidelity values as high as $(1-8.10^{-5})$ can be reached for a 100~$\mu$s long experiment. This fidelity decreases if the laser couplings are not sufficiently strong or if pulses are too short. Depending on the laser parameters, a detuning mismatch can be tolerated but relative frequency drifts of the lasers must be drastically avoided. If the laser pulses are interrupted before completion of the STIRAP-like process, a linear combination of  $D_{3/2}$ and $D_{5/2}$ can be built. If one of the lasers continues to be applied, the combination can also include the ground state  $S_{1/2}$. The cancellation of the first order Doppler effect by a geometric phase matching of the  laser beams allows the application of these methods to an ion cloud. On the contrary, like shown in \cite{champenois06}, if entanglement of internal and external degrees of freedom is needed, another geometry can be used which results in motional side-bands to the dark resonance and allows the process to include modification of the vibrational state along with transfer of the internal state.

\section*{Acknowledgement}
This work has been realized in the frame of the collaboration between Université des Sciences et Techniques de Masuku (USTM) and Aix-Marseille Université (AMU). M.R.K. would like to thank USTM and AMU for financial support during the completion of this work. T.B.E. acknowledges financial support from AMU and the European Physical Society (EPS). C.C. would like to thank A. Amy-Klein for precious information about relative frequency drift of optical combs.

\section*{References}
\bibliographystyle{unsrt}

\end{document}